\documentclass{article}
\newcommand{\be}{\begin{equation}}
\newcommand{\ee}{\end{equation}}
\newcommand{\nbar}[1]{\overline{#1}}

\def\bea{\begin{eqnarray}}
\def\eea{\end{eqnarray}}
\def\beas{\begin{eqnarray*}}
\def\eeas{\end{eqnarray*}}
\def\sla{\raise.15ex\hbox{$/$}\kern-.57em}

\def\parm{{\partial}^{-}}
\def\parp{\partial^+}

\newcommand{\delp}{{\partial^+}}
\newcommand{\del}{{\partial}}
\newcommand{\delb}{{\bar{\partial}}}
\newcommand{\phibar}{{\bar{\phi}}}
\newcommand{\qbar}{{\bar{q}_+}}
\newcommand{\qplus}{{q_+}}

\usepackage{amsmath}
\usepackage{latexsym}
\usepackage{graphicx}

\begin{document}
\begin{titlepage}
\begin{flushright}    AEI-2006-054 \\ 
\end{flushright}
\vskip 1cm
\centerline{\LARGE{\bf {The $\mathcal N=8$ Supergravity Hamiltonian}}}
\vskip .5cm
\centerline{\LARGE{\bf {as a Quadratic Form}}}

\vskip 1.5cm
\centerline{{Sudarshan Ananth$^\dagger$,\;\;Lars Brink$^*$,\;\;Rainer Heise$^\dagger$\;\;and\;\;Harald G. Svendsen$^\dagger$}} 
\vskip 1cm
\centerline{\em  $^\dagger$ Max-Planck-Institut f\"{u}r Gravitationsphysik}
\centerline{\em $\;$ Albert-Einstein-Institut, D-14476 Golm}
\vskip .5cm

\centerline{\em $^*$ Department of Fundamental Physics}
\centerline{\em $\;$ Chalmers University
of Technology, }
\centerline{\em S-412 96 G\"oteborg}

\vskip 1.5cm

\centerline{\bf {Abstract}}
\vskip .5cm
\noindent We conjecture that the light-cone Hamiltonian of $\mathcal N=8$ Supergravity can be expressed as a quadratic form. We explain why this rewriting is {\emph {unique}} to maximally supersymmetric theories. The $\mathcal N=8$ quartic interaction vertex is constructed and used to verify that this conjecture holds to order $\kappa^2$.
\vfill
\begin{flushleft}
July 2006 \\
\end{flushleft}
\end{titlepage}

\section{Introduction}
\vskip 1cm
\noindent Eleven-dimensional supergravity~\cite{CJS} on reduction to four dimensions yields the maximally supersymmetric $\mathcal N=8$ theory. $(\mathcal N=8,d=4)$ Supergravity is similar in many ways to the other maximally supersymmetric theory in four dimensions, $\mathcal N=4$ Yang-Mills. In reference~\cite{ABKR}, it was shown that the light-cone Hamiltonian of $\mathcal N=4$ Yang-Mills could be expressed as a quadratic form. The key to this rewriting was the maximal supersymmetry present in the theory. Since this is true in $\mathcal N=8$ as well, we conjecture that a similar rewriting must be possible. To start with, we show this explicitly to order $\kappa$. We explain why this quadratic-form structure is {\emph {unique}} to maximally supersymmetric theories and simply does not apply to other cases. 

\vskip 0.5cm

\noindent A light-cone superspace formulation of $\mathcal N=8$ Supergravity was first achieved in~\cite{BLN1,BBB2} wherein the authors constructed the action to order $\kappa$. The formulation has not been extended since. In this paper, we extend the action to order $\kappa^2$ by constructing the quartic interaction vertex. We will then see that the Hamiltonian is a quadratic form at order $\kappa^2$ as well. 

\vskip 0.5cm

\noindent When formulating a maximally supersymmetric theory (like $\mathcal N=8$ supergravity) in light-cone superspace, it suffices to show that the superspace action correctly reproduces the component action for any one component field (our focus will be on the graviton). Once this is done, the remaining component terms in the action will follow from supersymmetry transformations. We will therefore construct the quartic interaction vertex in $\mathcal N=8$ Supergravity by requiring that it correctly reproduce pure gravity in light-cone gauge. 

\vskip 0.5cm

\noindent The long-term aim of this program of research~\cite{ABKR,ABR2} is aimed at understanding the divergent behavior of $\mathcal N=8$ Supergravity. Curtright~\cite{TC} conjectured that any divergences occurring in the $\mathcal N=8$ theory, were attributible to the incomplete cancellation of Dynkin indices of $SO(9)$ representations. The light-cone offers an ideal framework to study this conjecture since it highlights the role of the spacetime little group. In addition, working on the light-cone ensures that we deal exclusively with the theory's physical degrees of freedom.

\vskip 0.5cm

\section{${\mathcal N}=8$ Supergravity in light-cone superspace}

\vskip 0.5cm

\noindent With the space-time metric $(-,+,+,+)$, the light-cone coordinates and their derivatives are 
\bea
\begin{split}
{x^{\pm}}&=&\frac{1}{\sqrt 2}\,(\,{x^0}\,{\pm}\,{x^3}\,)\ ;\qquad {\partial^{\pm}}=\frac{1}{\sqrt 2}\,(\,-\,{\partial_0}\,{\pm}\,{\partial_3}\,)\ , \\
x &=&\frac{1}{\sqrt 2}\,(\,{x_1}\,+\,i\,{x_2}\,)\ ;\qquad {\bar\partial} =\frac{1}{\sqrt 2}\,(\,{\partial_1}\,-\,i\,{\partial_2}\,)\ , \\
{\bar x}& =&\frac{1}{\sqrt 2}\,(\,{x_1}\,-\,i\,{x_2}\,)\ ;\qquad {\partial} =\frac{1}{\sqrt 2}\,(\,{\partial_1}\,+\,i\,{\partial_2}\,)\ .
\end{split}
\eea

\vskip 0.5cm

\noindent The degrees of freedom of $\mathcal N=8$ Supergravity theory may be captured in a single superfield~\cite{BLN1}. In terms of Grassmann variables $\theta^m$, which transform as the $8$ of $SU(8)$ ($m=1\,\ldots\,8$), we define\footnote{The factor of $4$ in the final term was missed in reference~\cite{BLN1}}
\bea
\begin{split}
\phi\,(\,y\,)\,=&\,\frac{1}{{\parp}^2}\,h\,(y)\,+\,i\,\theta^m\,\frac{1}{{\parp}^2}\,{\bar \psi}_m\,(y)\,+\,\frac{i}{2}\,\theta^m\,\theta^n\,\frac{1}{\parp}\,{\bar A}_{mn}\,(y)\ , \\
\;&-\,\frac{1}{3!}\,\theta^m\,\theta^n\,\theta^p\,\frac{1}{\parp}\,{\bar \chi}_{mnp}\,(y)\,-\,\frac{1}{4!}\,\theta^m\,\theta^n\,\theta^p\,\theta^q\,{\bar C}_{mnpq}\,(y)\ , \\
\;&+\,\frac{i}{5!}\,\theta^m\,\theta^n\,\theta^p\,\theta^q\,\theta^r\,\epsilon_{mnpqrstu}\,\chi^{stu}\,(y)\ ,\\
\;&+\,\frac{i}{6!}\,\theta^m\,\theta^n\,\theta^p\,\theta^q\,\theta^r\,\theta^s\,\epsilon_{mnpqrstu}\,\parp\,A^{tu}\,(y)\ ,\\
\,&+\,\frac{1}{7!}\,\theta^m\,\theta^n\,\theta^p\,\theta^q\,\theta^r\,\theta^s\,\theta^t\,\epsilon_{mnpqrstu}\,\parp\,\psi^u\,(y)\ ,\\
\,&+\,\frac{4}{8!}\,\theta^m\,\theta^n\,\theta^p\,\theta^q\,\theta^r\,\theta^s\,\theta^t\,\theta^u\,\epsilon_{mnpqrstu}\,{\parp}^2\,{\bar h}\,(y)\ ,
\end{split}
\eea

\noindent In this notation, the two-component graviton is represented by 
\be
\label{graviton}
h\,=\,\frac{1}{\sqrt 2}\,(\,h_{11}\,+\,i\,h_{12}\,)\ ;\qquad {\bar h}\,=\,\frac{1}{\sqrt 2}\,(\,h_{11}\,-\,i\,h_{12}\,)\ .
\ee
\noindent The ${\bar \psi}_m$ represent the spin-$\frac{3}{2}$ gravitinos in the theory. The $28$ gauge fields are ${\bar A}_{mn}$ and the corresponding gauginos are the ${\bar \chi}_{mnp}$. The ${\bar C}_{mnpq}$ are the $70$ scalar fields in the spectrum. Complex conjugation of the fields is denoted by a bar.

\vskip 0.5cm

\noindent All fields are local in the  modified light-cone coordinates  
\bea
y~=~\,(\,x,\,{\bar x},\,{x^+},\,y^-_{}\equiv {x^-}-\,\frac{i}{\sqrt 2}\,{\theta_{}^\alpha}\,{{\bar \theta}^{}_\alpha}\,)\ .
\eea

\noindent In this $LC_2$ (two-component light-cone) form, all the unphysical degrees of freedom have been integrated out. The superfield $\phi$ and its complex conjugate $\bar\phi$ satisfy chiral constraints,

\be
d^m\,\phi\,(\,y\,)\,=\,0\;\; ;\qquad {\bar d}_n\,{\bar \phi}\,(\,y\,)\,=\,0\ ,
\ee
\noindent where
\bea
d^{\,m}\,=\,-\,\frac{\partial}{\partial\,{\bar \theta}_m}\,-\,\frac{i}{\sqrt 2}\,\theta^m\,\parp\;\; ;\qquad{\bar d}_n\,=\,\frac{\partial}{\partial\,\theta^n}\,+\,\frac{i}{\sqrt 2}\,{\bar \theta}_n\,\parp\ ,
\eea

\noindent The superfield and its conjugate are furthermore related via the ``inside-out" constraint
\bea
\label{io}
\,{\phi}\,=\,\frac{1}{4}\,\frac{{(d\,)}^8}{{\parp}^4}\,{\bar \phi}\ ,
\eea
\noindent where ${(d\,)}^8\,=\,d^1\,d^2\,\ldots\,d^8$. This constraint is unique to maximally supersymmetric theories.

\vskip 0.5cm

\noindent On the light-cone, supersymmetry splits into two varieties~\cite{BBB2}, the kinematical ones
\bea
q^m_{\,+}\,=\,-\,\frac{\partial}{\partial\,{\bar \theta}_m}\,+\,\frac{i}{\sqrt 2}\,\theta^m\,\parp ;\qquad {\bar q}_{\,+\,n}=\;\;\;\frac{\partial}{\partial\,\theta^n}\,-\,\frac{i}{\sqrt 2}\,{\bar \theta}_n\,\parp\ ,
\eea
\noindent and the dynamical supersymmetries
\bea
\label{dynsus}
\begin{split}
q_-^{\,m}\,\equiv\,&i\,[\,\bar j^-\,,\,q^m_{\,+}\,]\,=\,\frac{\bar \partial}{\parp}\,q^m_{\,+}\,+\,{\cal O}(\kappa)\ , \\
{\bar q}_{-\,n}\,\equiv\,&i\,[\,j^-\,,\,{\bar q}_{\,+\,n}\,]\,=\,\frac{\partial}{\parp}\,{\bar q}_{\,+\,n}\,+\,{\cal O}(\kappa)\ .
\end{split}
\eea

\vskip 0.5cm

\subsection{The action to order $\kappa$}
\vskip 0.5cm

\noindent The ${\mathcal N}=8$ supergravity action to order $\kappa$ was derived using purely algebraic methods in reference~\cite{BBB2} and further simplified in reference~\cite{ABR2}. For a detailed listing of all the superPoincar\'e generators and their commutation relations, see references~\cite{ABKR,BBB2,ABR2}. The light-cone superspace action for $\mathcal N=8$ supergravity reads

\be
\label{n=8}
\beta\,\int\;d^4x\,\int d^8\theta\,d^8 \bar \theta\,{\cal L}\ ,
\ee
where $\beta\,=\,-\,\frac{1}{64}$ and
\bea
\label{one}
{\cal L}&=&-\bar\phi\,\frac{\Box}{\partial^{+4}}\,\phi\,-\,2\,\kappa\,(\,\frac{1}{{\parp}^2}\;{\nbar \phi}\;\;{\bar \partial}\,{\phi}\;{\bar \partial}\,{\phi}+\,\frac{1}{{\parp}^2}\;\phi\,\partial\,{\nbar \phi}\,\partial\,{\nbar \phi})\ .
\eea

\noindent The d'Alembertian is
\bea
\Box\,=\,2\,(\,\partial\,{\bar \partial}\,-\,\parp\,\parm\,)\ ,
\eea
\noindent $\kappa\,=\,{\sqrt {8\,\pi\,G}}$ and Grassmann integration is normalized such that $\int d^8\theta\,{(\theta)}^8=1$.
\vskip 0.5cm

\vskip 1cm

\section{The $\mathcal N=8$ Hamiltonian as a quadratic form}
\vskip 0.5cm
\noindent Maximally supersymmetric theories (like $\mathcal N=4$ Yang-Mills and $\mathcal N=8$ Supergravity) are special for various reasons. Specifically, the superfield governing these theories satisfies the inside-out constraint (equation (\ref {io})). This constraint allows us to express the Hamiltonian of the theory as a quadratic form. In this section, we illustrate this at lowest order and then at order $\kappa$.

\vskip 0.5cm
\subsection{Lowest order Hamiltonian}
\vskip 0.5cm

\noindent Based on comparison to $\mathcal N=4$ Yang-Mills~\cite{ABKR}, we conjecture that the light-cone Hamiltonian of $\mathcal N=8$ Supergravity can be expressed as
\bea
\label{claim}
{\cal H}~=~\frac{1}{4\,\sqrt{2}}\,(\,{\mathcal W}_{\,m}\,,\,{\mathcal W}_{\,m}\,)\ ,
\eea
with
\bea
{\mathcal W}_{\,m}\,=\,{\bar q}_{-\,m}\,\phi\ ,
\eea
and the inner product defined as
\be
\label{inner}
(\,\phi\,,\,\xi\,)~\equiv~2i\int d^4\!x\, d^8\theta\,d^8\,{\bar\theta}\;{\bar\phi}\,\frac{1}{{\parp}^3}\xi\ .
\ee

\noindent This relation (\ref{claim}), which we will prove presently (to order $\kappa^2$), is {\emph {unique}} to maximally supersymmetric theories. It is unrelated to the fact that the Hamiltonian is the anticommutator of two supersymmetries (this will become obvious in a few steps).

\vskip 0.5cm

\noindent We start by verifying this conjecture at the lowest order
\bea
\begin{split}
{\cal H}^0\,&=\,\frac{1}{4\,\sqrt{2}}\,(\,{\mathcal W}^0_m\,,\,{\mathcal W}^0_m\,)\ , \\
&=\,\frac{2i}{4\,\sqrt{2}}\,\int d^4\!x\, d^8\theta\,d^8\,{\bar\theta}\;\;q_-^{\,m}\,{\bar \phi}\,\frac{1}{{\parp}^3}\,{\bar q}_{-\,m}\,\phi\ ,
\end{split}
\eea

\noindent and rewrite this as two terms
\bea
\label{unik}
{\cal H}^0\,=\,\frac{i}{4\,\sqrt{2}}\,\,\int d^4\!x\, d^8\theta\,d^8\,{\bar\theta}\;\;{\Big (}\,q_-^{\,m}\,{\bar \phi}\,\frac{1}{{\parp}^3}\,{\bar q}_{-\,m}\,\phi\,+\,q_-^{\,m}\,{\bar \phi}\,\frac{1}{{\parp}^3}\,{\bar q}_{-\,m}\,\phi\,{\Big )}\ .
\eea

\vskip 0.5cm

\noindent In a non-maximally supersymmetric theory, this expression does not simplify further. 

\vskip 0.5cm

\subsubsection{Maximally supersymmetric theories}

\vskip 0.5cm

\noindent In the special case, where the supersymmetric theory under consideration is maximally supersymmetric, equation (\ref {unik}) can be further simplified. This is because such theories are always described by constrained superfields. Equation (\ref{io}) is the relevant constraint in the case of the $\mathcal N=8$ superfield. 

\vskip 0.5cm

\noindent We apply this constraint to the second term in equation (\ref {unik}) to obtain
\bea
{\cal H}^0\,=\,\frac{i}{4\,\sqrt{2}}\,\int d^4\!x\, d^8\theta\,d^8\,{\bar\theta}\;\;{\Big (}\,q_-^{\,m}\,{\bar \phi}\,\frac{1}{{\parp}^3}\,{\bar q}_{-\,m}\,\phi\,+\,\frac{1}{{\parp}^4}\,q_-^{\,m}\,\phi\,\parp\,{\bar q}_{-\,m}\,{\bar \phi}\,{\Big )}\ .
\eea

\noindent Using the explicit expressions for the dynamical supersymmetries from equation (\ref {dynsus}), we get
\bea
{\cal H}^0\,=\,\frac{i}{4\,\sqrt{2}}\,\,\int d^4\!x\, d^8\theta\,d^8\,{\bar\theta}\;\;{\Big (}\,\frac{\bar \partial}{\parp}\,q^m_{\,+}\,{\bar \phi}\,\frac{\partial}{{\parp}^4}\,{\bar q}_{\,+\,m}\,\phi\,+\,\frac{\bar \partial}{{\parp}^5}\,q^m_{\,+}\,\phi\,\partial\,{\bar q}_{\,+\,m}\,{\bar \phi}\,{\Big )}\ .
\eea
\noindent After some integration by parts, this may be reexpressed as
\bea
{\cal H}^0\,=\,\frac{i}{4\,\sqrt{2}}\,\int d^4\!x\, d^8\theta\,d^8\,{\bar\theta}\,\frac{\partial\,{\bar \partial}}{{\parp}^5}\,{\bar \phi}\,\{\,q^m_{\,+}\,,\,{\bar q}_{\,+\,m}\,\}\,\phi\ .
\eea
\noindent Since $\{\,q^m_{\,+}\,,\,{\bar q}_{\,+\,m}\,\}\,\phi\,=\,i\,8\,{\sqrt 2}\,\parp\,\phi$, we have
\bea
{\cal H}^0\,=\,\int{d^4}x\,{d^8}\theta\,{d^8}{\bar \theta}\,
{\bar \phi}\,\frac{2\,\partial\bar\partial}{{\parp}^4}\,\phi\ ,
\eea
\noindent which is indeed the correct kinetic term in the Hamiltonian describing the $\mathcal N=8$ theory~\cite{BBB2}.

\vskip 0.5cm
\subsection{Order $\kappa$}
\vskip 0.5cm

\noindent We now extend this to order $\kappa$. The dynamical supersymmetry generators are known to this order~\cite{BBB2} so we can write down the expressions for $\mathcal W$ and its complex conjugate,
\bea
\label{dublew}
{\mathcal W}_m\,=\,\frac{\partial}{\parp} {\bar q}_{+\,m}\,\phi\,+\,\kappa\,\frac{1}{\parp}\,{\Big (}\,{\bar \partial}\,{\bar d}_m\,\phi\,{{\parp}^2}\,\phi\,-\,\parp\,{\bar d}_m\,\phi\,\parp\,{\bar \partial}\,\phi\,{\Big )}\,+\,{\cal O}(\kappa^2)\ ,
\eea
\bea
\label{dublewbar}
{\nbar {\mathcal W}}^m\,=\,\frac{\bar \partial}{\parp}\,q_+^{\,m}\,{\bar \phi}\,+\,\kappa\,\frac{1}{\parp}\,{\Big (}\,\partial\,d^m\,{\bar \phi}\,{{\parp}^2}\,{\bar \phi}\,-\,\parp\,d^m\,{\bar \phi}\,\parp\,\partial\,{\bar \phi}\,{\Big )}\,+\,{\cal O}(\kappa^2)\ .
\eea

\vskip 0.5cm

\noindent With these, we directly compute the quadratic form
\bea
\frac{1}{4\,\sqrt{2}}\,(\,{\mathcal W}\,,\,{\mathcal W}\,)\,=\,\frac{2\,i}{4\,\sqrt{2}}\int d^4\!x\, d^8\theta\,d^8\,{\bar\theta}\;{\nbar {\mathcal W}}\,\frac{1}{{\parp}^3}\,{\mathcal W}\ .
\eea
\noindent The calculation is straightforward but fairly lengthy so details are relegated to Appendix {\bf {A}}. After simplification, the cubic interaction vertex occuring in the $\mathcal N=8$ Hamiltonian is
\bea
\int\;d^4x\,d^8\theta\,d^8{\bar \theta}\;\;2\,\kappa\,{\Big (}\,\frac{1}{{\parp}^2}\;{\nbar \phi}\;\;{\bar \partial}\,{\phi}\;{\bar \partial}\,{\phi}+\,\frac{1}{{\parp}^2}\;\phi\,\partial\,{\nbar \phi}\,\partial\,{\nbar \phi}{\Big )}\ ,
\eea

\noindent which is exactly equal to the equivalent term in the Lagrangian (obtained using different means) in equation (\ref {one}).

\vskip 1cm

\section{The Hamiltonian to order $\kappa^2$ : lessons from the quadratic form}

\vskip 0.5cm

\noindent The quadratic form studied so far will not immediately tell us the Hamiltonian to order $\kappa^2$. This is because the dynamical supersymmetry (and hence $\mathcal W$) is known only to order $\kappa$. However the quadratic form still offers a lot of insight into possible forms the quartic interaction may take.

\vskip 0.5cm

\noindent Our plan then is as follows. In this section, we will collect information from the quadratic form, dimensional analysis and helicity considerations. Based on these pointers, the general structure of the quartic interaction vertex will become much clearer. We will then make a guess at such an interaction vertex and check that the guess is correct. 

\vskip 0.5cm

\noindent In order to check our Ansatz, we will set all terms of the superfield (in the superspace vertex) except the graviton to zero. The resulting vertex (in components) must reproduce pure gravity in the light-cone gauge. Once the graviton vertex is fixed, the supersymmetry transformations (which are manifest within the superfield) will produce the remaining components of the action.

\vskip 0.5cm

\subsection{Information from the quadratic form}
\vskip 0.5cm

\noindent With this general plan in mind, we first draw information from the fact that the Hamiltonian is expressible as a quadratic form. Notice that the $\mathcal N=8$ Hamiltonian (at order $\kappa^2$) has two contributions from the quadratic form. The first is
\bea
(\,{\mathcal W}^\kappa\,,\,{\mathcal W}^\kappa\,)\ .
\eea

\noindent This is straightforward to compute and reads
\bea
\label{kapkap}
\begin{split}
-\,\frac{i}{2\,\sqrt 2}\,\kappa^2\,\int d^4\!x\, d^8\theta\,d^8\,{\bar\theta}\,\frac{1}{\delp^5}\,&{\Big (}\,\delp^2\phi\,{\bar d}\,\delb\,\phi\,-\,\delp{\bar d}\,\phi\,\delp\delb\,\phi\,{\Big )} \\
\times\,&{\Big (}\,\delp^2\phibar\,d\,\del\,\phibar\,-\,\delp d\,\phibar\,\delp\del\,\phibar\,{\Big )}.
\end{split}
\eea
\noindent The second contribution is
\bea
\label{info}
(\,{\mathcal W}^0\,,\,{\mathcal W}^{\kappa^2}\,)\,+\,(\,{\mathcal W}^{\kappa^2}\,,\,{\mathcal W}^0\,)\ .
\eea
\noindent Even though ${\mathcal W}^{\kappa^2}$ is unknown, there is still structural information in this expression. The first of the two terms in (\ref {info}) is
\bea
(\,{\mathcal W}^0\,,\,{\mathcal W}^{\kappa^2}\,)\,=\,2i\int d^4\!x\, d^8\theta\,d^8\,{\bar\theta}\;\frac{\bar \partial}{{\parp}^4}\,q^+\,{\bar \phi}\,{\mathcal W}^{\kappa^2}\ ,
\eea
\noindent where the measure factor $\frac{1}{{\parp}^3}$ has been integrated on to the ${\mathcal W}^0$ term. This tells us that the quartic interaction has the form
\bea
\label{form}
\kappa^2\,\frac{\bar \partial}{{\parp}^4}\,q^+\,{\bar \phi}\,\cdot\,\aleph
\eea
\noindent We know (from comparison to equation (\ref {kapkap})) that $\aleph$ involves two chiral superfields and one anti-chiral superfield.

\vskip 0.5cm

\subsection{Dimensional analysis and helicity considerations}

\vskip 0.5cm

\noindent We list below the helicities and length-dimensions (with $\hbar\,=\,1$) of some quantities at our disposal. 

\begin{center}
\begin{tabular}{||c c c||}
\hline\hline
Variable & Helicity ($h$) & Dimension ($D$) \\
\hline
$\phi$ & $+2$ & $+1$  \\
$\bar \phi$ & $-2$ & $+1$  \\
$\partial$ & $+1$ & $-1$ \\
${\bar \partial}$ & $-1$ & $-1$ \\
$\parp$ & $0$ & $-1$ \\
$d^m\,,\,q^m$ & $+\,1/2$ & $-\,1/2$  \\
${\bar d}_n\,,\,{\bar q}_n$ & $-\,1/2$ & $-\,1/2$ \\
\hline\hline
\end{tabular}
\end{center}
\vskip .5cm

\noindent The quartic interaction term at order $\kappa^2$ presently reads (up to a constant),
\bea
\int\;d^4x\,d^8\theta\,d^8{\bar \theta}\,\kappa^2\,(\,\frac{\bar \partial}{{\parp}^4}\,q^+\,{\bar \phi}\,)\,\cdot\,\aleph\ .
\eea

\noindent The measure has dimension $-\,4$, $\kappa^2$ has dimension $+\,2$. It is then easy to see that $\aleph$ has dimension $-\,\frac{3}{2}$ and helicity $\frac{5}{2}$. 

\vskip 0.5cm

\noindent From equation (\ref {kapkap}) and the table we see that $\aleph$ must contain: two $\phi$s, one $\bar \phi$, one $\bar \partial$ and either a $\bar d$ or a $\bar q$. The dimensions can be taken care of by introducing the necessary number of $\parp$s and $\frac{1}{\parp}$s.

\vskip 0.5cm

\noindent {\bf {The issue of chiralization}}

\vskip 0.5cm

\noindent ${\mathcal W}^{\kappa^2}$ must commute with both $d$ and $\bar d$. This is necessary to ensure that the dynamical supersymmetry generator respects the chirality of the superfield it acts on. Thus
\bea
{\mathcal W}^{\kappa^2}\,=\,{\cal C}(\aleph)\ ,
\eea
\noindent where $\cal C$ represents chiralization and is explained in Appendix {\bf {B}}. In the quartic vertex, this appears as 
\bea
\int\;\frac{1}{{\parp}^3}\,{\nbar {\mathcal W}}^0\,{\mathcal W}^{\kappa^2}\,=\,\kappa^2\,\int\;\frac{1}{{\parp}^3}\,{\nbar {\mathcal W}}^0\,\cdot\,{\biggl [}\,\aleph\,+\,{{\bar d}_{m_1\,\ldots\,m_a}}\,\{\,\ldots\,\}\,{\biggr ]}\ .
\eea
\noindent However, the $\bar d\,$s that appear in the chiralized expression (on partial integration) annihilate the ${\nbar {\mathcal W}}^0$ (since it is anti-chiral). Hence the chiralizing terms are not relevant to the calculation at this order.

\vskip 0.5cm

\subsection{The Ansatz for the quartic interaction}

\vskip 0.5cm

\noindent Our Ansatz for the quartic interaction vertex is simply $(\,{\mathcal W}^\kappa\,,\,{\mathcal W}^\kappa\,)$ plus
\bea
\label{genans}
\begin{split}
&c^A_{abcd}\;\frac{1}{{\parp}^a}\,{\biggl (}\,{\parp}^b\,\phi\;{\parp}^c\,{\bar q}_+\,\phi\,{\biggr )}\,{\parp}^d\,\partial\,{\bar \phi}\;\cdot\;\frac{1}{{\parp}^3}\,{\nbar {\cal W}}^0 \\
+\;\;&c^B_{abcd}\;\frac{1}{{\parp}^a}\,{\biggl (}\,{\parp}^b\,{\bar q}_+\,\phi\;{\parp}^c\partial\,\phi\,{\biggr )}\,{\parp}^d\,{\bar \phi}\;\cdot\;\frac{1}{{\parp}^3}\,{\nbar {\cal W}}^0 \\
+\;\;&c^C_{abcd}\;\frac{1}{{\parp}^a}\,{\biggl (}\,{\parp}^b\,\phi\;{\parp}^c\,{\bar q}_+\,\partial\,\phi\,{\biggr )}\,{\parp}^d\,{\bar \phi}\;\cdot\;\frac{1}{{\parp}^3}\,{\nbar {\cal W}}^0\ ,
\end{split}
\eea
\noindent subject to the constraint
\bea
\begin{split}
a\,+\,3\,&=\,b\,+\,c\,+\,d\ .
\end{split}
\eea
\vskip 0.5cm
\noindent The constants $c^A_{abcd}$, $c^B_{abcd}$ and $c^C_{abcd}$ are to be fixed by comparison with the gravity action.

\vskip 0.5cm

\noindent It is such a general Ansatz that we expand using Mathematica (details in Appendix {\bf {D}}). We only focus on the graviton components and require that the resulting expression reproduce the quartic vertex for pure gravity in the light-cone gauge. 

\vskip 0.5cm

\noindent Gravity on the light-cone has been studied previously by numerous authors~\cite{SS,MK,GS,BCL,AK}. A brief but self contained review of the formalism is presented in Appendix {\bf {C}}.

\vskip 1cm

\section{The $\mathcal N=8$ Supergravity action to order $\kappa^2$}
\vskip 0.5cm
\noindent By explicit comparison with the gravity vertex, we find that the $\mathcal N=8$ Supergravity quartic interaction vertex is
\bea
\begin{split}
-\,\kappa^2\,\frac{i}{\sqrt 2}\,{\biggl [}\;\;X\,+\,{\nbar X}\,+\,\frac{1}{2}\,\frac{1}{\delp^5}\,&{\Big (}\delp^2\phi\,{\bar d}\,\delb\,\phi\,-\,\delp{\bar d}\,\phi\,\delp\delb\,\phi{\Big )} \\
\times\,&{\Big (}\delp^2\phibar\,d\,\del\,\phibar\,-\,\delp d\,\phibar\,\delp\del\,\phibar{\Big )}\;\;{\biggr ]}, \\
\end{split}
\eea
\noindent where the explicit form of $X$ is given in Appendix {\bf {D}}. The final answer (as explained in the appendix) confirms our conjecture regarding the quadratic form, to order $\kappa^2$.

\vskip 0.5cm

\noindent The answer does not seem to simplify very much based on the usual tricks (partial integrations, inside-out constraints and so on). Our result is surprisingly lengthy for a superspace expression, however it is important to remember that the entire component action for the $\mathcal N=8$ theory in light-cone gauge involves many thousand terms. Also, unlike $\mathcal N=4$ Yang-Mills, where the dynamical supersymmetry generator stops at order $\kappa$, the $\mathcal N=8$ supersymmetry generator extends to all orders. 

\vskip 0.5cm

\noindent It is worth noting that there is no freedom arising from superfield redefinitions. This is because any shift of the form
\bea
\phi\;\rightarrow\;\phi\,+\,\phi^3
\eea
\noindent in the kinetic term will result in time-derivatives $\parm$ reappearing in the action to order $\kappa^2$. These time derivatives (as explained in the gravity section) have been pushed to higher orders and no longer occur at $\kappa^2$.

\vskip 1cm

\section{Summary}
\noindent The quartic interaction governing $\mathcal N=8$ supergravity has been constructed. In addition, we have shown (to order $\kappa^2$) that the Hamiltonian is expressible as a quadratic form. We believe this quadratic structure of the Hamiltonian will hold to higher orders as well. We expect to return to this issue and explore its consequences further.

\vskip 0.5cm

\noindent The general structure of the four-point interaction term is quite simple. The complication stems from the many ways the various derivatives enter. However, when computing the Feynman rules, momentum conservation will help us to reduce the number of terms, especially if we use some specific frame for the momenta. This should lead to tractable forms for 
the simpler diagrams. The fact remains that supergraph computations, essential to understanding the divergent nature of this theory, are going to prove extremely tedious. It would be very satisfying if the quartic interaction could be recast in an elegant manner. The light-cone formalism was essential in the proof of finiteness of $\mathcal N=4$ Yang-Mills~\cite{BLN2} and one hope is that the techniques used in that paper will prove equally useful when studying $\mathcal N=8$ Supergravity.

\vskip 0.5cm

\noindent Finally, it will be interesting to see if the quartic interaction simplifies considerably on reduction to three dimensions. Such a simplification is likely to occur due to the large $SO(16)$ symmetry that the $d=3$ theory is known to possess~\cite{HN}.

\vskip 1cm

\noindent {\bf {Acknowledgments}}

\vskip 0.3cm

\noindent We are grateful to Pierre Ramond for many valuable discussions. We thank Stefano Kovacs, Hermann Nicolai, G. Rajasekaran and Hidehiko Shimada for helpful comments. HGS is supported by a Humboldt fellowship.

\newpage

\renewcommand{\theequation}{A-\arabic{equation}}
\setcounter{equation}{0}  
\section*{Appendix {\bf {A}}}  

\vskip 0.5cm
\subsection*{Detailed computation of the Hamiltonian to order $\kappa$}

\vskip 1cm

\noindent In this section, we drop the spinor indices for convenience. Start by considering 
\bea
\label{fir}
{\cal H}\,=\,\frac{i}{2\,{\sqrt 2}}\,\int\,{\nbar {\mathcal W}}\,\frac{1}{{\parp}^3}\,{\mathcal W}\ .
\eea
\noindent We will not carry the constant $\frac{i}{2\,{\sqrt 2}}$ any further. Instead we will reintroduce it into the expression at the end. Substituting the expression for ${\mathcal W}^\kappa$ from equation (\ref{dublew}) we obtain
\bea
{\cal H}^\kappa\,=\,\kappa\,\int\,\frac{\bar \partial}{{\parp}^5}\,{q_+}\,{\bar \phi}\,\{{\bar \partial}\,{\bar d}\,\phi\,{{\parp}^2}\,\phi\,-\,\parp\,{\bar d}\,\phi\,\parp\,{\bar \partial}\,\phi\,\}\,+\,{\mbox {c.c}}\ .
\eea
\noindent We will ignore the complex conjugate in this section. For any chiral combination $X$, ${q_+}\,X\,=\,i\,{\sqrt 2}\,\parp\,\theta\,X$. This yields 
\bea
\label{appx}
\begin{split}
{\cal H}^\kappa\,&=\,i\,{\sqrt 2}\,\kappa\,\int\,\frac{\bar \partial}{{\parp}^4}\,\theta\,{\bar \phi}\,\{{\bar \partial}\,{\bar d}\,\phi\,{{\parp}^2}\,\phi\,-\,\parp\,{\bar d}\,\phi\,\parp\,{\bar \partial}\,\phi\,\}\ , \\
&=\hskip 3.5cm\,Y\,\hskip 1cm\,+\,\hskip 1cm\,X
\end{split}
\eea

\noindent We now focus on $Y$ and rewrite it using the fact that $\theta\,{\bar d}\,=\,8\,-\,{\bar d}\,\theta$ to obtain
\bea
\begin{split}
Y\,&=\,-\,i\,8\,{\sqrt 2}\,\kappa\,\int\,\frac{\bar \partial}{{\parp}^4}\,{\bar \phi}\,{\bar \partial}\,\phi\,{{\parp}^2}\,\phi\,-\,i\,{\sqrt 2}\,\kappa\,\int\,\frac{\bar \partial}{{\parp}^4}\,{\bar \phi}\,\theta\,{\bar \partial}\,\phi\,{{\parp}^2}\,{\bar d}\,\phi\ , \\
&=\hskip 3.5cm\,A\,\hskip 1cm\,+\,\hskip 1cm\,B
\end{split}
\eea

\noindent Now partially integrate $B$ with respect to $\parp$ to get
\bea
\begin{split}
B\,&=\,-\,i\,{\sqrt 2}\,\kappa\,\int\,\frac{\bar \partial}{{\parp}^3}\,{\bar \phi}\,\,{\bar \partial}\,\phi\,\parp\,\theta\,{\bar d}\,\phi\,-\,X\ , \\
&=\hskip 2cm\,G\,\hskip 1cm\,-\,\hskip 1cm\,X
\end{split}
\eea
\noindent and we simply cancel the second term against the $X$ that appears in equation (\ref {appx}).

\noindent We now focus on term $G$. To simplify calculations, rather than track $\theta\,\cdot\,{\bar d}$, we will focus on $\theta^1\,{\bar d}_1$ for the moment and then multiply by a factor of $8$ at the end (although we are missing this factor of $8$, we will continue to refer to term $G$ as $G$). In term $G$, we now impose the inside-out constraint (equation (\ref {io})) on the middle superfield to obtain
\bea
G\,=\,-\,i\,{\sqrt 2}\,\kappa\,\int\,\frac{\bar \partial}{{\parp}^3}\,{\bar \phi}\,\,{\bar \partial}\,\frac{d^{1\ldots8}}{4\,{{\parp}^4}}\,{\bar \phi}\,\parp\,\theta^1\,{\bar d}_1\,\phi\,
\eea
\noindent We then integrate the chiral derivatives away from the middle superfield to obtain
\bea
\label{modf}
\begin{split}
G\,=\,&-\,i\,{\sqrt 2}\,\kappa\,\int\,{\bar \partial}\parp\,\phi\,\,\frac{\bar \partial}{{\parp}^4}\,{\bar \phi}\,\parp\,\theta^1\,{\bar d}_1\,\phi\, \\
&+\,i\,{\sqrt 2}\,\kappa\,\int\,\frac{\bar \partial}{{\parp}^3}\,{\bar \phi}\,\,{\bar \partial}\,\frac{d^{2\ldots8}}{4\,{{\parp}^4}}\,{\bar \phi}\,\parp\,\theta^1\,(\,-\,i\,{\sqrt 2}\,\parp\,)\,\phi\, \\
=&\hskip 1cm\,H\, \\
&+\,\hskip 0.5cm\,I
\end{split}
\eea
\noindent Term $I$ has two $\parp$s acting on the last superfield so we integrate one of these out and obtain
\bea
\begin{split}
I\,&=\,i\,{\sqrt 2}\,\kappa\,\int\,{\bar \partial}\parp\,{\bar d}_1\,\phi\,\,\frac{\bar \partial}{{\parp}^4}\,{\bar \phi}\,\parp\,\theta^1\,\phi\,+\,i\,{\sqrt 2}\,\kappa\,\int\,{\bar \partial}\,{\bar d}_1\,\phi\,\,\frac{\bar \partial}{{\parp}^3}\,{\bar \phi}\,\parp\,\theta^1\,\phi\, \\
&=\hskip 2cm\,J\,\hskip 2cm\,+\,\hskip 2cm\,K
\end{split}
\eea
\noindent Term $J$ is rewritten as
\bea
\begin{split}
J\,&=\,-\,i\,{\sqrt 2}\,\kappa\,\int\,{\bar \partial}\parp\,\phi\,\,\frac{\bar \partial}{{\parp}^4}\,{\bar \phi}\,\parp\,\phi\,-\,H\ , \\
&=\hskip 2cm\,L\,\hskip 0.5cm\,-\,\hskip 0.5cm\,H\ ,
\end{split}
\eea
\noindent with the second term canceling against the $H$ that occurs in equation (\ref {modf}).

\noindent Now term $K$ also simplifies to
\bea
\begin{split}
K\,&=\,-\,i\,{\sqrt 2}\,\kappa\,\int\,{\bar \partial}\,\phi\,\,\frac{\bar \partial}{{\parp}^3}\,{\bar \phi}\,\parp\,\phi\,-\,G\ .
\end{split}
\eea

\noindent This leads to (after putting back the factor of 8)
\bea
G\,=\,i\,4\,{\sqrt 2}\,\kappa\,\int\,{\bar \partial}\,\phi\,\,\frac{\bar \partial}{{\parp}^4}\,{\bar \phi}\,{\parp}^2\,\phi\ .
\eea
\noindent After taking into account, the constant from equation (\ref {fir}), we obtain the final answer,
\bea
{\cal H}^\kappa\,=\,A\,+\,G\,=\,2\,\kappa\,\int\,{\bar \partial}\,\phi\,\,\frac{\bar \partial}{{\parp}^4}\,{\bar \phi}\,{\parp}^2\,\phi\ .
\eea
\noindent Imposing the inside-out constraint on the third superfield yields
\bea
{\cal H}^\kappa\,=\,2\,\kappa\,\int\,\frac{1}{{\parp}^2}\,{\bar \phi}\,{\bar \partial}\,\phi\,{\bar \partial}\,\phi\ ,
\eea
\noindent and proves the quadratic form structure of the Hamiltonian at order $\kappa$.

\newpage

\renewcommand{\theequation}{B-\arabic{equation}}
\setcounter{equation}{0}  
\section*{Appendix {\bf {B}}}  
\vskip 0.5cm

\subsection*{Chiralization}
\vskip 0.5cm

\noindent The object $\aleph$ is simply the dynamical supersymmetry generator at order $\kappa^2$. All generators must respect the chirality of the superfields they act on and hence must commute with both $d^m$ and ${\bar d}_m$. $\aleph$ contains two chiral superfields and one anti-chiral superfield. In addition it contains either a ${\bar d}$ or a ${\bar q}$. We choose the ${\bar q}$ since it commutes with both chiral derivatives.

\vskip 0.5cm

\noindent Now, $\aleph$ needs to be chiralized. This can be accomplished through a descent procedure. There seem to be two non-equivalent ways of doing this.

\noindent For any general non-chiral expression of the form, $A\,{\bar B}$ (where $A$ is any compound chiral function and $\bar B$ a compound anti-chiral function), we define a ``chiral product"

\subsubsection*{Scheme I}

\bea
{\cal C}\,(\,A{\bar B}\,)\,=\,A{\bar B}\,+\,{\sum_{a=1}^8}\,{\frac {{(-1)}^a}{a!}}\,{\frac {{\bar d}_{m_1\,\ldots\,m_a}}{{(-\,i\,{\sqrt 2}\,\parp)}^a}}\;(\,A\,{d^{m_a\,\ldots\,m_1}}\,{\bar B}\,)\ ,
\eea

\noindent or 

\subsubsection*{Scheme II}

\bea
{\cal C}\,(\,A{\bar B}\,)\,=A{\bar B}\,+\,{\sum_{a=1}^8}\,{\frac {{(-1)}^a}{a!}}\,{\frac {{\bar d}_{m_1\,\ldots\,m_a}}{{(-\,i\,{\sqrt 2}\,\parp)}^a}}\,A\,{d^{m_a\,\ldots\,m_1}}\,{\bar B}\ .
\eea

\noindent in both cases,
\bea
{\bar d}_{m_1\,\ldots\,m_a}\,=\,{{\bar d}_{m_1}}\,\ldots\,{{\bar d}_{m_a}}\ ;\qquad {d^{m_a\,\ldots\,m_1}}\,=\,d^{m_a}\,\ldots\,d^{m_1}\ .
\eea
\vskip 0.3cm
\noindent ${\cal C}\,(\,A{\bar B}\,)\,$ via either scheme is now a chiral function and satisfies $d\,{\cal C}\,=\,0$.

\vskip 0.5cm

\noindent Notice that the quartic interaction from equation (\ref {form}) is of the form
\bea
\kappa^2\,\frac{\bar \partial}{{\parp}^4}\,q^+\,{\bar \phi}\,\cdot\,\aleph\ .
\eea 

\noindent Adopting Scheme {\bf {I}} for chiralizing $\aleph$ is more convenient because all the ``chiralizing" terms simply vanish. This is because $\aleph$ is multiplied by an antichiral field. The relevant expression is

\bea
\kappa^2\,\frac{\bar \partial}{{\parp}^4}\,q^+\,{\bar \phi}\,\cdot\,{\biggl [}\,\aleph\,+\,{{\bar d}_{m_1\,\ldots\,m_a}}\,\{\,\ldots\,\}\,{\biggr ]}\ .
\eea 

\noindent A simple partial integration of the overall $\bar d$ onto the antichiral superfield on the left, kills all the chiralizing terms. Hence {\bf {at this order}}, chiralization of $\aleph$ is unimportant. 

\vskip 0.5cm

\noindent It is possible that Scheme {\bf {II}} may yield more interesting results as far as the quartic interaction is concerned.

\newpage

\renewcommand{\theequation}{C-\arabic{equation}}
\setcounter{equation}{0}  
\section*{Appendix {\bf {C}}}  

\subsection*{Einstein gravity in the $LC_2$ formalism}
\vskip 0.5cm

\noindent This section offers a quick review of pure gravity in light-cone gauge. This section follows closely, the treatment in reference~\cite{BCL}. Other references dealing with light-cone gravity include~\cite{SS,MK,GS,AK}.
\vskip 0.5cm

\noindent The Einstein-Hilbert action reads
\bea
S_{EH}=\int\,{d^4}x\,L\,=\,\frac{1}{2\,\kappa^2}\,\int\,{d^4}x\,{\sqrt {-g}}\,R\ ,
\eea
\noindent where $g=\det{g_{\mu\nu}}$ and $R$ is the curvature scalar.

\vskip 0.5cm

\subsubsection*{Gauge choices}

\vskip 0.5cm

\noindent Light-cone gauge is chosen by setting
\bea
g_{--}\,=\,g_{-i}\,=\,0\ .
\eea
\noindent A fourth gauge choice will be made shortly. The metric is parametrized as follows
\bea
\begin{split}
g_{+-}\,&=\,-\,e^\phi\ , \\
g_{i\,j}\,&=\,e^\psi\,\gamma_{i\,j}\ .
\end{split}
\eea
\noindent The fields $\phi\,,\,\psi$ are real while $\gamma_{ij}$ is a $2\,\times\,2$ real, symmetric, unimodular matrix. 

\vskip 0.5cm

\noindent We us the equations of motion
\bea
R_{\mu\nu}\,=\,0\ ,
\eea
\noindent to eliminate the unphysical degrees of freedom. The $R_{-i}\,=\,0$ relation implies that 
\bea
\begin{split}
g^{-i}\,=\,-\mathrm{e}^{-\,\phi}\,\frac{1}{\partial^+}\bigg[\,&\,\gamma^{ij}\,\mathrm{e}^{\phi\,-\,2\,\psi}\,\frac{1}{\parp}\,{\Big \{}\,\mathrm{e}^{-\,\psi}\,{\Big (}\,\frac{1}{2}\,\parp\,\gamma^{kl}\,\partial_j\,\gamma_{kl}\,-\,\parp\,\partial_j\,\phi \\
&\,-\,\parp\,\partial_j\,\psi\,+\,\partial_j\phi\,\parp\,\psi\,{\Big )}\,+\,\partial_l\,{\Big (}\,\mathrm{e}^{-\,\psi}\,\gamma^{kl}\,\parp\,\gamma_{jk}\,{\Big )}\,{\Big \}}\,\bigg]\ ,
\end{split}
\eea
\noindent while $R_{--}=0$ gives 
\bea
\label{con2}
2\,\parp\,\phi\,\parp\,\psi\,-\,2\,{\parp}^2\,\psi\,-\,{(\parp\,\psi)}^2\,+\,\frac{1}{2}\,\parp\,\gamma^{kl}\,\parp\,\gamma_{kl}\,=\,0\ .
\eea
\noindent We solve equation (\ref{con2}) by choosing
\bea
\phi\,=\,\frac{1}{2}\,\psi\ .
\eea
\noindent This constitutes our fourth gauge choice. Equation (\ref {con2}) yields
\bea
\psi=\frac{1}{4} \frac{1}{{\partial^+}^2} \bigg[ \partial^+ \gamma^{i j} \partial^+ \gamma_{i j}\bigg].
\eea

\vskip 0.5cm

\subsubsection*{The action in light cone gauge}

\vskip 0.5cm

\noindent The $LC_2$ Lagrangian for gravity is
\bea
L=\frac{1}{2 \kappa^2} \sqrt{-g} \left( 2 g^{+-} R_{+-} +g^{i j} R_{i j}\right)\ .
\eea
\noindent Based on the metric choices, this becomes
\bea
  L&=&\frac{1}{2 \kappa^2} \Bigg(
  \mathrm{e}^{\psi} 
  \bigg( \partial^-\partial^+ \phi + \partial^+ \partial^- \psi
  -\frac{1}{2}\partial^- \gamma^{ij} \partial^+ \gamma_{ij} \bigg)\nonumber\\&&
   -\mathrm{e}^{\phi} \gamma^{i j}
  \bigg( \partial_i\partial_j \phi +\frac{1}{2} \partial_i \phi
  \partial_j \phi - \partial_i \phi \partial_j \psi - \frac{1}{4}
  \partial_i\gamma^{k l} \partial_j \gamma_{k l} 
  + \frac{1}{2}\partial_i \gamma^{k l} \partial_k \gamma_{j l} \bigg)
  \nonumber \\ &&
  - \frac{1}{2}\mathrm{e}^{\phi -2\psi} \gamma^{i j}
  \frac{1}{{\partial^+}}R_i \frac{1}{{\partial^+}}R_j 
      \Bigg),
      \label{finallag}
\eea
\noindent where 
\bea
R_i&=&  \mathrm{e}^{\psi} 
\bigg( \frac{1}{2}\partial^+ \gamma^{j k} \partial_i \gamma_{j k} -
  \partial^+ \partial_i \phi -\partial_i
  \partial^+ \psi + \partial_i \phi \partial^+ \psi \bigg) \nonumber\\&&
  + \partial_k \bigg( \mathrm{e}^{\psi}\gamma^{j k} \partial^+
  \gamma_{ij} \bigg).
  \label{req}
\eea

\vskip 0.5cm

\subsubsection*{Perturbative expansion}

\vskip 0.5cm
 
\noindent In order to obtain a perturbative expansion of the metric, we set 

\begin{eqnarray}
\gamma_{i j}&=&\left(\mathrm{e}^{\kappa H}\right)_{i j}\ , \\
H&=&\begin{pmatrix} h_{11} & h_{12}\\h_{12} &-h_{11}\end{pmatrix}\ .
\eea

\noindent The physical graviton is represented by $h$ and $\bar h$ which were defined in equation (\ref {graviton}). In terms of these fields,
\bea
\psi=-\frac{1}{{\partial^+}^2} 
\bigg[ \partial^+ h \partial^+ \bar h\bigg]+{\cal O}(h^4)\ .
\eea

\noindent At this order, we ignore terms of order ${\cal O}(h^4)$ and beyond in $\psi$. The Lagrangian (\ref {finallag}) is then
\begin{eqnarray}
  \!\!L\!\!\!&\!\!=\!\!&\frac{1}{2} h\, \Box \, \bar h\\
  &&+\kappa\, \bar h \,{\partial^+}^2 \bigg[ -h \frac{{\bar
  \partial}^2}{{\partial^+}^2} h + \frac{{\bar
  \partial}}{{\partial^+}} h \frac{\bar
  \partial}{\partial^+} h \bigg] +c.c.\\
  && + \kappa^2 \Bigg[ \frac{1}{{\partial^+}^2} \left[ \partial^+ h 
\partial^+ \bar h \right]\,  \frac{\partial \bar \partial + 
2\,\partial^- \partial ^+ }{{\partial^+}^2} \left[ \partial^+ h 
\partial^+ \bar h \right] \nonumber \\
  && \phantom{+}
   -\frac{1}{{\partial^+}^2} \left[ \partial^+ h \partial^+ \bar h 
\right]\, \left( \partial^- h \partial ^+ \bar h +\partial^+ h
  \partial ^- \bar h\right) \nonumber \\
  &&\phantom{+ }
    + \frac{1}{3} \partial^-\bar h\left( h\,\bar h \partial^+h - h^2
  \partial^+ \bar h \right) + \frac{1}{3} \partial^- h\left( h\,\bar h
  \partial^+ \bar h - {\bar h}^2   \partial^+ h \right) \nonumber \\
  &&\phantom{+}
   -\frac{1}{{\partial^+}^2} \left[ \partial^+ h \partial^+ \bar h 
\right]\,
    \left(2 \partial \bar \partial h \, \bar h+ 2 h \partial \bar
  \partial \bar h + 9 \bar \partial h \partial \bar  h + \partial h
  \bar \partial \bar h
  \right. \nonumber \\
  &&\phantom{+ }
    \left.
    - 2 \frac{\partial \bar \partial}{\partial^+} h \, \partial^+\bar h
  - 2 \partial^+ h \frac{\partial \bar \partial}{\partial^+} \bar h  
  + 3 \frac{1}{\partial^+} \left[ \partial \bar
  \partial h \, \partial^+\bar h+ \partial^+ h \partial \bar
  \partial \bar h \right] \right) \nonumber \\
  &&\phantom{+ }
 - h\,\bar h\,\left(\frac{4}{3} \partial \bar \partial h \, \bar
  h + \frac{4}{3} h \partial \bar  
  \partial \bar h + 2\, \bar \partial h \partial \bar  h   
  +4 \frac{\partial \bar \partial}{\partial^+} h \, \partial^+\bar h
  +4 \partial^+ h \frac{\partial \bar \partial}{\partial^+} \bar h  
  \right)\nonumber \\
  &&\phantom{+}
  -2 \frac{1}{\partial^+} \left[
    2 \, \bar \partial h\, \partial ^+\bar h  
    + h \partial^+ \bar \partial \bar h  
    - \partial^+ \bar \partial h
  \bar h \right]  
  \,h\,\partial \bar h \nonumber \\
&&\phantom{+ }
  -2 \frac{1}{\partial^+} \left[
    2 \,\partial^+ h\, \partial \bar h
    + \partial^+  \partial h \, \bar h
    - h \partial^+ \partial \bar h \right] \,\bar \partial h \, \bar h
  \\
  &&\phantom{+}
  -\frac{1}{\partial^+}\! \left[
    2 \bar \partial h\, \partial ^+\bar h  
    + h \partial^+ \bar \partial \bar h  
    - \partial^+ \bar \partial h
  \bar h \right]
  \frac{1}{\partial^+}\! \left[
    2 \partial^+ h\, \partial \bar h
    + \partial^+  \partial h \, \bar h
    - h \partial^+ \partial \bar h \right]
  \Bigg]\ . \nonumber
\end{eqnarray}
\noindent We then shift all the light-cone time-derivatives $\parm$ that occur in this action to higher orders in $\kappa$. This is achieved by means of the field redefinition
\begin{eqnarray}
  h &\to& h - \kappa^2 \frac{1}{\partial^+}\left[ 2\,
  {\partial^+}^2h \, \frac{1}{{\partial^+}^3} \left[ \partial^+ h 
\partial^+ \bar h \right] + \partial^+ h \,\frac{1}{{\partial^+}^2} 
\left[ \partial^+ h \partial^+ \bar h \right] \nonumber \right.  \\
    &&\phantom{ h - \kappa^2 \frac{1}{\partial^+}}  \left.
 +   \frac{1}{3} \left( h\,\bar h \partial^+h - h^2
  \partial^+ \bar h \right)\right]\ .
\end{eqnarray}

\vskip 0.5cm

\subsubsection*{The shifted Lagrangian}
\vskip 0.5cm

\noindent After the shift, the $LC_2$ Lagrangian for pure gravity reads
\begin{eqnarray}
 \!\!L^{\kappa^2}\!\!\!&\!=\!&\!\!\frac{1}{{\partial^+}^2} 
    \bigg[ \partial^+ h \partial^+ \bar h\bigg]\frac{\partial \bar \partial}{{\partial^+}^2} 
    \bigg[ \partial^+ h \partial^+ \bar h\bigg]\nonumber \\
  && + \frac{1}{{\partial^+}^3} \bigg[ \partial^+ h \partial^+ \bar h\bigg] \left( \partial \bar
  \partial h \, \partial^+\bar h+ \partial^+ h \partial \bar
  \partial \bar h \right) \nonumber\\
&&
  -\frac{1}{{\partial^+}^2} 
    \bigg[ \partial^+ h \partial^+ \bar h\bigg]\, \left(2\, \partial \bar \partial h \, \bar h+ 2\, h \partial \bar
  \partial \bar h + 9\, \bar \partial h \partial \bar  h + \partial h
  \bar \partial \bar h \right. \nonumber\\
&& \phantom{-\frac{1}{{\partial^+}^2} \bigg[ \partial^+ h \partial^+ \bar h\bigg]\, } 
	\left.
  -  \frac{\partial \bar \partial}{\parp} h \, \partial^+\bar h
  -  \partial^+ h \frac{\partial \bar \partial}{\partial^+} \bar h 
  \right)
   \nonumber  \\
  &&
  -2 \frac{1}{\partial^+} \left[
    2 \bar \partial h\, \partial ^+\bar h 
    + h \partial^+ \bar \partial \bar h 
    - \partial^+ \bar \partial h
  \bar h \right] 
  \,h\,\partial \bar h \nonumber \\
&&
  -2 \frac{1}{\partial^+} \left[
    2 \partial^+ h\, \partial \bar h
    + \partial^+  \partial h \, \bar h
    - h \partial^+ \partial \bar h \right] \,\bar \partial h \, \bar h\nonumber 
  \\
  &&
  -\frac{1}{\partial^+} \left[
    2 \bar \partial h\, \partial ^+\bar h 
    + h \partial^+ \bar \partial \bar h 
    - \partial^+ \bar \partial h
  \bar h \right]
  \frac{1}{\partial^+} \left[
    2 \partial^+ h\, \partial \bar h
    + \partial^+  \partial h \, \bar h
    - h \partial^+ \partial \bar h \right] \nonumber\\
  && - h\,\bar h\,\left(\partial \bar \partial h \, \bar  h + h
  \partial \bar  
  \partial \bar h + 2\, \bar \partial h \partial \bar  h 
  +3 \frac{\partial \bar \partial}{\partial^+} h \, \partial^+\bar h
  +3 \partial^+ h \frac{\partial \bar \partial}{\partial^+} \bar h 
  \right)\ .
\end{eqnarray}
\vskip 0.5cm

\noindent Notice that the shift is unique. Any additional shift in $h$ will reintroduce the $\parm$ (from the kinetic term) into the action.

\newpage

\renewcommand{\theequation}{D-\arabic{equation}}
\setcounter{equation}{0}  
\section*{Appendix {\bf {D}}}  

\vskip 0.5cm

\subsection*{The quartic interaction}

\vskip 0.5cm

\noindent Assuming that our conjecture is valid to order $\kappa^2$, the form of $X$ (from Section {\bf {5}}) ought to be

\bea
\int\,{\nbar {\cal W}}^0\,\frac{1}{{\parp}^3}\,{\cal W}^{\kappa^2}\ .
\eea
\noindent This is indeed what we find. The explicit expression for $X$ is shown below and proves that the conjecture holds to this order. The Mathematica code used in these computations is available for download from this site:

\vskip 0.5cm

\noindent {\bf {http://www.aei.mpg.de/$\sim$harald/supergravity.html}}

\vskip 0.5cm

\begin{equation}
X\,=\,L_A\,+\,L_B\,+\,L_C\ .
\end{equation}
\noindent We use $A$, $B$ and $C$ to denote the three classes of terms that occur in the general ansatz (\ref {genans}). Subscripts $_1$ and $_2$ differentiate between terms that have an overall $\frac{1}{{\parp}^3}$ and terms that have an overall $\frac{1}{\parp}$ (acting on the first two superfields).
\vskip 0.5cm

\begin{align}
   L_{A,1}=
   & \frac{3}{4}~\frac{1}{\delp^3}\Bigl(
  \delp\phi~\delp\qbar\phi\Bigr) 
  \delp^4\del\phibar~\frac{1}{\delp^4}\qplus\delb\phibar
\end{align}
\begin{align}
  L_{A,2}=
    \frac{53}{12}& ~\frac{1}{\delp}\Bigl(
  \frac{1}{\delp}\phi~\frac{1}{\delp}\qbar\phi \Bigr)~ 
  \delp^6\del\phibar~\frac{1}{\delp^4}\qplus\delb\phibar
\\
    +\frac{265}{12}& ~\frac{1}{\delp}\Bigl(
  \frac{1}{\delp}\phi~\qbar\phi \Bigr)~ 
  \delp^5\del\phibar~\frac{1}{\delp^4}\qplus\delb\phibar
\\
    +\frac{265}{6}& ~\frac{1}{\delp}\Bigl(
  \frac{1}{\delp}\phi~\delp\qbar\phi \Bigr)~ 
  \delp^4\del\phibar~\frac{1}{\delp^4}\qplus\delb\phibar
\\
    +\frac{265}{6}& ~\frac{1}{\delp}\Bigl(
  \frac{1}{\delp}\phi~\delp^2\qbar\phi \Bigr)~ 
  \delp^3\del\phibar~\frac{1}{\delp^4}\qplus\delb\phibar
\\
    +\frac{265}{12}& ~\frac{1}{\delp}\Bigl(
  \frac{1}{\delp}\phi~\delp^3\qbar\phi \Bigr)~ 
  \delp^2\del\phibar~\frac{1}{\delp^4}\qplus\delb\phibar
\\
    +\frac{53}{12}& ~\frac{1}{\delp}\Bigl(
  \frac{1}{\delp}\phi~\delp^4\qbar\phi \Bigr)~ 
  \delp\del\phibar~\frac{1}{\delp^4}\qplus\delb\phibar
\\
    +\frac{119}{12}& ~\frac{1}{\delp}\Bigl(
  \phi~\frac{1}{\delp}\qbar\phi \Bigr)~ 
  \delp^5\del\phibar~\frac{1}{\delp^4}\qplus\delb\phibar
\\
    +\frac{358}{9}& ~\frac{1}{\delp}\Bigl(
  \phi~\qbar\phi \Bigr)~ 
  \delp^4\del\phibar~\frac{1}{\delp^4}\qplus\delb\phibar
\\
   +\frac{553}{9}& ~\frac{1}{\delp}\Bigl(
  \phi~\delp\qbar\phi \Bigr)~ 
  \delp^3\del\phibar~\frac{1}{\delp^4}\qplus\delb\phibar
\\
    +\frac{121}{3}& ~\frac{1}{\delp}\Bigl(
  \phi~\delp^2\qbar\phi \Bigr)~ 
  \delp^2\del\phibar~\frac{1}{\delp^4}\qplus\delb\phibar
\\
    +\frac{373}{36}& ~\frac{1}{\delp}\Bigl(
  \phi~\delp^3\qbar\phi \Bigr)~ 
  \delp\del\phibar~\frac{1}{\delp^4}\qplus\delb\phibar
\\
    +\frac{1}{9}& ~\frac{1}{\delp}\Bigl(
  \phi~\delp^4\qbar\phi \Bigr)~ 
  \del\phibar~\frac{1}{\delp^4}\qplus\delb\phibar
\\
    +\frac{193}{36}& ~\frac{1}{\delp}\Bigl(
  \delp\phi~\frac{1}{\delp}\qbar\phi \Bigr)~ 
  \delp^4\del\phibar~\frac{1}{\delp^4}\qplus\delb\phibar
\\
    +\frac{305}{18}& ~\frac{1}{\delp}\Bigl(
  \delp\phi~\qbar\phi \Bigr)~ 
  \delp^3\del\phibar~\frac{1}{\delp^4}\qplus\delb\phibar
\\
    +\frac{67}{4}& ~\frac{1}{\delp}\Bigl(
  \delp\phi~\delp\qbar\phi \Bigr)~ 
  \delp^2\del\phibar~\frac{1}{\delp^4}\qplus\delb\phibar
\\
    +\frac{107}{18}& ~\frac{1}{\delp}\Bigl(
  \delp\phi~\delp^2\qbar\phi \Bigr)~ 
  \delp\del\phibar~\frac{1}{\delp^4}\qplus\delb\phibar
\\
    +\frac{1}{9}& ~\frac{1}{\delp}\Bigl(
  \delp\phi~\delp^3\qbar\phi \Bigr)~ 
  \del\phibar~\frac{1}{\delp^4}\qplus\delb\phibar
\end{align}

\begin{align}
  L_{B,1}=
    -\frac{3}{2}& ~\frac{1}{\delp^3}\Bigl(
  \delp\qbar\phi~\del\phi\Bigr) 
  \delp^5\phibar~\frac{1}{\delp^4}\qplus\delb\phibar
\\
    -& \frac{1}{\delp^3}\Bigl(
  \qbar\phi ~\delp\del\phi\Bigr) 
  \delp^5\phibar~\frac{1}{\delp^4}\qplus\delb\phibar
\\
    -\frac{1}{2}& ~\frac{1}{\delp^3}\Bigl(
  \delp\qbar\phi ~\delp\del\phi \Bigr) 
  \delp^4\phibar~\frac{1}{\delp^4}\qplus\delb\phibar
\\
    -& \frac{1}{\delp^3}\Bigl(
  \delp^2\qbar\phi ~\delp\del\phi \Bigr) 
  \delp^3\phibar~\frac{1}{\delp^4}\qplus\delb\phibar
\\
    -& \frac{1}{\delp^3}\Bigl(
  \delp\qbar\phi ~\delp^2\del\phi \Bigr) 
  \delp^3\phibar~\frac{1}{\delp^4}\qplus\delb\phibar
\end{align}
\begin{align}
L_{B,2}=
    -\frac{1}{2}& ~\frac{1}{\delp}\Bigl(
  \frac{1}{\delp}\qbar\phi~\delp^2\del\phi  \Bigr) 
  \delp^3\phibar~\frac{1}{\delp^4}\qplus\delb\phibar
\\
   -\frac{1}{2}&~\frac{1}{\delp}\Bigl(
  \frac{1}{\delp}\qbar\phi~\delp^3\del\phi \Bigr) 
  \delp^2\phibar~\frac{1}{\delp^4}\qplus\delb\phibar
\\
    +\frac{3}{2}&~\frac{1}{\delp}\Bigl(
   \qbar\phi~\delp\del\phi \Bigr) 
  \delp^3\phibar~\frac{1}{\delp^4}\qplus\delb\phibar
\\
    -\frac{1}{2}&~\frac{1}{\delp}\Bigl(
   \qbar\phi~\delp^2\del\phi \Bigr) 
  \delp^2\phibar~\frac{1}{\delp^4}\qplus\delb\phibar
\\
    -\frac{1}{2}&~\frac{1}{\delp}\Bigl(
   \delp\qbar\phi~\delp\del\phi \Bigr) 
  \delp^2\phibar~\frac{1}{\delp^4}\qplus\delb\phibar
\\
    +&~\frac{1}{\delp}\Bigl(
  \frac{1}{\delp}\qbar\phi~\del\phi \Bigr) 
  \delp^5\phibar~\frac{1}{\delp^4}\qplus\delb\phibar
\end{align}

\begin{align}
  L_{C,1}=
    -\frac{3}{2}& ~\frac{1}{\delp^3}\Bigl(
  \phi~\delp\qbar\del\phi \Bigr) 
  \delp^5\phibar~\frac{1}{\delp^4}\qplus\delb\phibar
\\
    -\frac{11}{4}& ~\frac{1}{\delp^3}\Bigl(
  \delp\phi~\del\qbar\phi \Bigr) 
  \delp^5\phibar~\frac{1}{\delp^4}\qplus\delb\phibar
\\
    -\frac{1}{2}& ~\frac{1}{\delp^3}\Bigl(
  \delp\phi~\delp\qbar\del\phi \Bigr) 
  \delp^4\phibar~\frac{1}{\delp^4}\qplus\delb\phibar
\\
    -& ~\frac{1}{\delp^3}\Bigl(
  \delp\phi~\delp^2\qbar\del\phi \Bigr) 
  \delp^3\phibar~\frac{1}{\delp^4}\qplus\delb\phibar
\\
    -& ~\frac{1}{\delp^3}\Bigl(
  \delp^2\phi~\delp\qbar\del\phi \Bigr) 
  \delp^3\phibar~\frac{1}{\delp^4}\qplus\delb\phibar
\end{align}
\begin{align}
  L_{C,2}=
    -\frac{1}{6} &~\frac{1}{\delp}\Bigl(
  \delp\phi~\delp\qbar\del\phi \Bigr) 
  \delp^2\phibar~\frac{1}{\delp^4}\qplus\delb\phibar
\\
    -\frac{13}{6}& ~\frac{1}{\delp}\Bigl(
  \delp^2\phi~\frac{1}{\delp}\qbar\del\phi \Bigr) 
  \delp^3\phibar~\frac{1}{\delp^4}\qplus\delb\phibar
\\
    +\frac{1}{6}& ~\frac{1}{\delp}\Bigl(
  \delp^2\phi~\qbar\del\phi \Bigr) 
  \delp^2\phibar~\frac{1}{\delp^4}\qplus\delb\phibar
\\
    -\frac{1}{6}& ~\frac{1}{\delp}\Bigl(
  \delp^2\phi~\delp\qbar\del\phi \Bigr) 
  \delp\phibar~\frac{1}{\delp^4}\qplus\delb\phibar
\\
    -\frac{179}{18}& ~\frac{1}{\delp}\Bigl(
  \delp^3\phi~\frac{1}{\delp}\qbar\del\phi \Bigr) 
  \delp^2\phibar~\frac{1}{\delp^4}\qplus\delb\phibar
\\
    -\frac{407}{18}& ~\frac{1}{\delp}\Bigl(
  \delp^3\phi~\qbar\del\phi \Bigr) 
  \delp\phibar~\frac{1}{\delp^4}\qplus\delb\phibar
\\
    -\frac{199}{12}& ~\frac{1}{\delp}\Bigl(
  \delp^3\phi~\delp\qbar\del\phi \Bigr) 
  \phibar~\frac{1}{\delp^4}\qplus\delb\phibar
\\
    -\frac{53}{12}& ~\frac{1}{\delp}\Bigl(
  \delp^3\phi~\delp^2\qbar\del\phi \Bigr) 
  \frac{1}{\delp}\phibar~\frac{1}{\delp^4}\qplus\delb\phibar
\\
    -\frac{191}{18}& ~\frac{1}{\delp}\Bigl(
  \delp^4\phi~\frac{1}{\delp}\qbar\del\phi \Bigr) 
  \delp\phibar~\frac{1}{\delp^4}\qplus\delb\phibar
\\
    -\frac{221}{18}& ~\frac{1}{\delp}\Bigl(
  \delp^4\phi~\qbar\del\phi\Bigr) 
  \phibar~\frac{1}{\delp^4}\qplus\delb\phibar
\\
    -\frac{53}{12}& ~\frac{1}{\delp}\Bigl(
  \delp^4\phi~\delp\qbar\del\phi\Bigr) 
  \frac{1}{\delp}\phibar~\frac{1}{\delp^4}\qplus\delb\phibar
\end{align}

\end{document}